\documentclass{article}[12pt, a4paper]

\usepackage{amsmath,amsthm, amssymb}

\makeatletter
\renewcommand\section{\@startsection 
{section}{1}{0pt}{-3.5ex plus-1ex minus-.2ex} 
{2.3ex plus.2ex}{\normalfont\large\bfseries}} 
\let\@ldthebibliography\thebibliography
\renewcommand{\thebibliography}[1]{
\centering
        \@ldthebibliography{#1}
}
\makeatother

\newtheoremstyle{theorem}%name
{10pt} % space above
{10pt} % space below
{\sl} % bofy font
{\parindent} % ident - empty=no indent, \parindent= paragraph indent
{\bf} % thm head font
{. } % punctuation after thm head
{ } % space after thm head: ` `=normal \newline=linebreak
{} % thm head specification
\theoremstyle{theorem}
\newtheorem{theorem}{Theorem}
\newtheorem*{theorem*}{Theorem}
\newtheorem*{lemma*}{Lemma}

\newtheoremstyle{defi}%name
{10pt} % space above
{10pt} % space below
{\rm} % bofy font
{\parindent} % ident - empty=no indent, \parindent= paragraph indent
{\bf} % thm head font
{. } % punctuation after thm head
{ } % space after thm head: ` `=normal \newline=linebreak
{} % thm head specification
\theoremstyle{defi}

\begin{document}

\title{\bf\large EMP AND LINEAR SCHR\"ODINGER MODELS FOR A CONFORMALLY BIANCHI I COSMOLOGY}
\author{\normalsize Jennie D'Ambroise\\
\normalsize Department of Mathematics and Statistics\\
\normalsize University of
 Massachusetts\\
\normalsize Amherst, MA 01003, USA\\
\normalsize e-mail:  dambroise@math.umass.edu\\}
\date{}
\maketitle
\thispagestyle{empty}

\noindent{\bf Abstract:}
Using a conformal version of the Bianchi I metric and a perfect fluid energy-momentum tensor, we show that the resulting Einstein field equations are equivalent to a generalized Ermakov-Milne-Pinney equation.  Using a transformation analogous to the known correspondence between the classical EMP and a linear Schr\"odinger equation, we then construct the linear Schr\"odinger formulation for this cosmological model.  Also, a note is included to show the extra term that arises when the energy-momentum tensor is taken to include a second term in addition to the scalar field.\\
{\bf AMS Subj. Classification:}  83C05, 83C15

%\begin{equation}T^{(2)}=-\left(\begin{array}{cccc}[abc]^2\rho&0&0&0\\0&a^2p&0&0\\0&0&b^2p&0\\0&0&0&c^2p\\\end{array}.\right)\end{equation}  

\section{\protect\centering Introduction}
\setcounter{equation}{0}
A number of recent papers have explicated the Ermakov-Milne-Pinney (EMP) and Schr\"odinger content of various scalar field cosmologies, including both 2+1 and 3+1 curved FRLW \cite{3,4, 5,6} and some anisotropic Bianchi models\cite{1,2,7}.  These connections are interesting since such equations often appear in condensed matter physics as well as quantum field theory; that is, such reformulations open the door to directly relating gravitational and nongravitational systems.  For the conformally Bianchi I metric that we will consider here, F. L. Williams \cite{7} has explicated a transformation which reduces the system of Einstein's field equations to a generalized EMP equation whose coefficient functions are coupled to a second equation.  Here we show a similar transformation which results in an uncoupled generalized EMP equation.  By implementing a transformation analogous to that which is known to relate the classical EMP to a linear Schr\"odinger equation, we further reduce the field equations to a linear Schr\"odinger equation.  The author thanks Floyd L. Williams for his support and advice on the results presented here.

\section{\protect\centering Einstein Equations}
\setcounter{equation}{0}
Consider the Einstein field equations $G_{ij}=K^2T_{ij}$ for metric 
\begin{equation}ds^2=-[a(t)b(t)c(t)]^2dt^2+a(t)^2dx^2+b(t)^2dy^2+c(t)^2dz^2\end{equation}
where $a(t), b(t), c(t) > 0$.  The perfect fluid energy-momentum tensor  is taken to be $T_{ij}=-\phi_{;i}\phi_{;j}+g_{ij}\left[\frac{1}{2}g^{k\lambda}\phi_{;k}\phi_{;\lambda}+V\circ\phi\right]$ for some scalar field $\phi$ with potential $V$.  Under these assumptions the field equations become
\begin{eqnarray}
\frac{\dot{a}\dot{b}}{ab}+\frac{\dot{a}\dot{c}}{ac}+\frac{\dot{b}\dot{c}}{bc} 
\stackrel{(i)}{=} 
K^2\left[\frac{\dot{\phi}^2}{2}+(abc)^2\left(V\circ\phi\right) \right]\\
\frac{\ddot{b}}{b}-\frac{\dot{b}^2}{b^2}+\frac{\ddot{c}}{c}-\frac{\dot{c}^2}{c^2}-\frac{\dot{a}\dot{b}}{ab}-\frac{\dot{a}\dot{c}}{ac}-\frac{\dot{b}\dot{c}}{bc} 
\stackrel{(ii)}{=} 
K^2\left[-\frac{\dot{\phi}^2}{2}+(abc)^2\left(V\circ\phi\right) \right]\notag\\
\frac{\ddot{a}}{a}-\frac{\dot{a}^2}{a^2}+\frac{\ddot{c}}{c}-\frac{\dot{c}^2}{c^2}-\frac{\dot{a}\dot{b}}{ab}-\frac{\dot{a}\dot{c}}{ac}-\frac{\dot{b}\dot{c}}{bc} \stackrel{(iii)}{=} 
K^2\left[-\frac{\dot{\phi}^2}{2}+(abc)^2\left(V\circ\phi\right) \right]\notag\\
\frac{\ddot{a}}{a}-\frac{\dot{a}^2}{a^2}+\frac{\ddot{b}}{b}-\frac{\dot{b}^2}{b^2}-\frac{\dot{a}\dot{b}}{ab}-\frac{\dot{a}\dot{c}}{ac}-\frac{\dot{b}\dot{c}}{bc} \stackrel{(iv)}{=} 
K^2\left[-\frac{\dot{\phi}^2}{2}+(abc)^2\left(V\circ\phi\right) \right]\notag
 \end{eqnarray}
where $K^2=8\pi G$ and $G$ is Newton's constant. 
\setlength{\headheight}{1.46cm}

\section{\protect\centering EMP Formulation}
\setcounter{equation}{0}

As in \cite{7}, equating pairs of the left sides of (ii)-(iv) in (2.2), allows us to deduce that  
\begin{equation}b(t)=C_1e^{\lambda t}a(t), \qquad c(t)=C_2e^{\mu t}a(t)\end{equation}
for real numbers $C_1,C_2 >0$ and $\lambda, \mu$.  Forming the combination $3(i)-(ii)-(iii)-(iv)$ and using the relations (3.1), we obtain
\begin{equation}4\frac{\dot{a}^2}{a^2}-\frac{\ddot{a}}{a}+2(\lambda+\mu)\frac{\dot{a}}{a}+\lambda\mu=\frac{K^2}{2}\dot{\phi}^2.\end{equation}
Introducing the quantity $\beta(t)=e^{(\lambda+\mu)t}a(t)^3$, equation $(3.2)$ becomes 
\begin{equation}2\frac{\dot{\beta}^2}{\beta^2}-\frac{\ddot{\beta}}{\beta}-(\lambda^2-\lambda\mu+\mu^2)=\frac{3}{2}K^2\dot{\phi}^2.\end{equation}
Now find some function $\overline\tau(t)$ such that 
\begin{equation}\dot{\overline\tau}(t)=\theta \beta(t)^{(6+n)/6}\end{equation}
 for any $n\neq 0$, $\theta>0$ and define $\phi_1(\tau)=\phi(\overline{f}(\tau))$ where $\overline{f}(\tau)$ is the inverse of $\overline\tau(t)$.  Then by (3.3), one can show that 
\begin{equation}y(\tau)=\beta(\overline{f}(\tau))^{n/6}\qquad\mbox{ and }\qquad Q(\tau)=\frac{nK^2}{4}\phi_1 ' (\tau)^2\end{equation}
satisfy the generalized EMP
\begin{equation}y''(\tau)+Q(\tau)y(\tau)=\frac{-n(\lambda^2-\lambda\mu+\mu^2)}{6\theta^2y(\tau)^{(12+n)/n}}.\end{equation}
Similarly we may state the converse implication, which produces a solution of the Einstein equations (2.2) from a solution of the generalized EMP (3.6).
\newpage
\begin{theorem}
Suppose that $y(\tau)$ is a solution of the generalized EMP (3.6) for a known function $Q(\tau)$ and for some constants $n\neq 0$, $\theta>0$ and $\lambda,\mu$.  Then a solution $(a, b, c, \phi ,V)$ of (2.2) can be constructed as follows.  First find functions $\overline\tau(t)$ and $\phi_1(\tau)$ such that
\begin{equation}
\dot{\overline\tau}(t) =\theta y(\overline\tau(t))^{(6+n)/n}\,, \quad \phi_1 '(\tau)^2 =\frac{4}{nK^2}Q(\tau).
\end{equation}
Then the following solve the Einstein equations (i)-(iv) in (2.2):
\begin{equation}a(t)=y(\overline\tau(t))^{2/n}e^{-\frac{1}{3}(\lambda+\mu)t},\end{equation}
\begin{equation}b(t)=C_1e^{\lambda t}a(t), \quad c(t)=C_2e^{\mu t}a(t), \quad \phi(t)=\phi_1(\overline\tau(t)),\end{equation}
\begin{equation} V\circ\phi=\frac{\theta^2}{nK^2C_1C_2}\left[\frac{12}{n}(y')^2-\frac{n(\lambda^2-\lambda\mu+\mu^2)}{3\theta^2(y)^{12/n}}-2(y)^2Q\right]\circ\overline\tau\end{equation}
for any constants $C_1,C_2>0$.
\end{theorem}

In the case $n=6$, (3.6) reduces to the classical EMP equation, which is known to be equivalent to a linear Schr\"odinger equation.  A similar transformation relates the generalized EMP in (3.6) to the linear Schr\"odinger equation 
\begin{equation}u''(x)+[E-P(x)]u(x)=0.\end{equation}
This can be shown by letting $u(x)=y(\tau(g(x)))^{-6/n}$ where $g(x)$ is the inverse of $\sigma(t)$ which satisfies $\dot{\sigma}(t)=1/\dot{\tau}(t)$, and where $\tau(t)$ is such that $\dot{\tau}(t)=\theta^{1/2}y(\tau(t))^{(6+n)/2n}$.  Then $u(x)$ satisfies (3.11) for $E=-(\lambda^2-\lambda\mu+\mu^2)$ and $P(x)=\frac{6\theta^2}{n}Q(\tau(g(x)))y(\tau(g(x)))^{(12+2n)/n}$.  This motivates the next section, in which we directly relate Einstein's field equations (2.2) to the linear Schr\"odinger equation (3.11).

\section{\protect\centering Linear Schr\"odinger Formulation}
\setcounter{equation}{0}

Beginning with a solution quintet $(a,b,c,\phi,V)$ of the Einstein equations (i)-(iv) in (2.2), one can follow the steps outlined above and obtain the formulas (3.1) for $b(t)$ and $c(t)$ in terms of $a(t)$ for some constants $C_1, C_2>0$ and $\lambda, \mu$.  Again, we deduce (3.2) and (3.3) for $\beta(t)=e^{(\lambda+\mu)t}a(t)^3$, find $\overline\tau(t)$ solving (3.4) for any $n\neq 0$ and $\theta>0$, and define $\phi_1(\tau)=\phi(\overline{f}(\tau))$ where $\overline{f}(\tau)$ is the inverse of $\overline{\tau}(t)$.  Now to obtain the Schr\"odinger formulation, we additionally solve for both $\tau(t)$ and $\sigma(t)$ that satisfy
\begin{equation}\dot{\tau}(t)=\dot{\overline{\tau}}(\overline{f}(\tau(t)))^{1/2},\qquad \dot{\sigma}(t)=\frac{1}{\dot{\tau}(t)},\end{equation}
and we denote by $g(x)$ the inverse function of $\sigma(t)$.  Introducing the quantities
\begin{equation}u(x)=\beta(\overline{f}(\tau(g(x)))^{-1}\qquad\mbox{ and }\qquad \psi(x)=\phi(\overline{f}(\tau(g(x)))),\end{equation}
one can show that $u(x)$, $P(x)=\frac{3K^2}{2}\psi ' (x)^2$ and $E=-(\lambda^2-\lambda\mu+\mu^2)$ solve the linear Schr\"odinger equation (3.11).  We may now state the converse implication.
\newpage
\begin{theorem}
Suppose that $u(x)$ is a solution of the linear Schr\"odinger equation (3.11) for some known function $P(x)$ and constant $E<0$.  First find functions $\sigma(t)$ and $\tau(t)$ such that 
\begin{equation}\dot{\sigma}(t)=u(\sigma(t))^{(6+n)/12} \quad \mbox{and} \quad \dot{\tau}(t)=\frac{1}{\dot{\sigma}(t)}\end{equation}
for some $n\neq 0$, and let $f\equiv \tau^{-1}$.  Next find $\psi(x)$ such that $\psi '(x)^2=\frac{2}{3K^2}P(x)$ and let $\lambda,\mu$ be any two constants such that 
\begin{equation}-(\lambda^2-\lambda\mu+\mu^2)=E.\end{equation}
Finally, solve for $\overline{\tau}(t)$ such that
\begin{equation}\dot{\overline{\tau}}(t)=\dot{\tau}(f(\overline{\tau}(t)))^2.\end{equation}
Then the following functions solve the Einstein equations (i)-(iv) in (2.2):
\begin{equation}a(t)=\left[e^{(\lambda+\mu)t}u(\sigma(f(\overline\tau(t))))\right]^{-1/3},\end{equation}
\begin{equation}b(t)=C_1e^{\lambda t} a(t), \quad c(t)=C_2e^{\mu t} a(t), \quad \phi(t)=\psi(\sigma(f(\overline{\tau}(t)))),\end{equation}
\begin{equation}V\circ\phi=\frac{1}{3K^2C_1^2C_2^2}\left[(u')^2+[E-P]u^2\right]\circ \sigma\circ f \circ \overline\tau.\end{equation}
\end{theorem}

We note that including a second term in the energy-momentum tensor $T_{ij}$ that has density $\rho(t)$ and pressure $p(t)$, one obtains a corresponding quantity
\begin{equation}-\frac{nK^2}{4\theta^2}C_1^2C_2^2\frac{(\rho( \overline{f}(\tau))+p(\overline{f}(\tau)))}{y(\tau)}\end{equation}
on the right side of the generalized EMP (3.6).  Similarly (3.11) becomes nonlinear with the term
\begin{equation}\frac{3K^2C_1^2C_2^2(\overline{\rho}(x)+\overline{p}(x))}{2u(x)}\end{equation}
on the right side of the equation where $\overline{\rho}(x)\equiv\rho(\overline{f}(\tau(g(x))))$ and 
$\overline{p}(x)\equiv p(\overline{f}(\tau(g(x))))$.  In this case both Theorems 1 and 2 include an additional $-\rho(t)$ added to the definitions of $V\circ\phi$.  

\section{\protect\centering Examples}
\setcounter{equation}{0}

Take the solution $u(x)=Ae^{-\sqrt{-E}x}$ to the linear Schr\"odinger equation (3.11) for $A >0$ with $E<0$ and $P(x)=0$.  Solving the differential equations (4.3) for $\sigma(t)$ and $\tau(t)$ in the case where $n=6$, we obtain
\begin{equation}\sigma(t)=\frac{1}{\sqrt{-E}}\ln\left[A\sqrt{-E}t\right]\qquad\mbox{ and }\qquad\tau(t)=\frac{\sqrt{-E}}{2}t^2.\end{equation}
In this case the inverse of $\tau(t)$ is $f(\tau)=\sqrt{\frac{2}{\sqrt{-E}}\tau}$ and so equation (4.5) becomes $\dot{\overline\tau}(t)=2\sqrt{-E}\overline\tau(t)$ 
therefore $\overline\tau(t)=c_0e^{2\sqrt{-E}t}$ for any $c_0>0$.  Using these functions and (4.6)-(4.7), we obtain
\newpage
\begin{equation}a(t)=\left(2c_0\sqrt{-E}\right)^{1/6}e^{\frac{1}{3}(\sqrt{-E}-\lambda-\mu)t}\end{equation}
\begin{equation}b(t)=C_1\left(2c_0\sqrt{-E}\right)^{1/6}e^{\frac{1}{3}(\sqrt{-E}+2\lambda-\mu)t}\end{equation}
\begin{equation}c(t)=C_2\left(2c_0\sqrt{-E}\right)^{1/6}e^{\frac{1}{3}(\sqrt{-E}-\lambda+2\mu)t}
\end{equation}
for $\lambda,\mu$ as in (4.4).  Since $P(x)=0$, $\psi(x)=\psi _0\equiv$ any constant therefore $\phi(t)=\psi_0$ also.  Since $\phi$ is constant, $V$ is also constant by (4.8) which then reduces to the formula
\begin{equation}V=\frac{1}{3K^2C_1C_2}\left[(u')^2+Eu^2\right].\end{equation}
For this example $V=0$, so that (5.2)-(5.4) is a vacuum solution.  Note that one can show by contradiction that for any nonzero $\lambda, \mu$ satisfying (4.4) then $\sqrt{-E}-\lambda-\mu<0$, $\sqrt{-E}+2\lambda-\mu>0$ and $\sqrt{-E}-\lambda+2\mu>0$. 

As another example, take $u(x)=Ae^{-\sqrt{-E}x}-\frac{1}{4A\sqrt{-E}}e^{\sqrt{-E}x}$ for any real number $A$ and any $E<0$.  Again, take $P(x)=0$.  Solving (4.3) for $n=6$, we obtain
\begin{equation}\sigma(t)=\frac{1}{2\sqrt{-E}}\ln\left[4A^2\sqrt{-E}\tanh^2\left(\frac{\sqrt[4]{-E}}{2}t\right)\right]\quad\mbox{ and }\quad\tau(t)=\cosh(\sqrt[4]{-E}t).\end{equation}
Using that $\dot{\tau}(f(\tau))=1/f'(\tau)=\sqrt[4]{-E}\sqrt{\tau^2-1}$, which is obtained by differentiating the relation $f(\tau(t))=t$ with respect to $t$, equation (4.5) becomes $\dot{\overline\tau}(t)=\sqrt{-E}(\overline\tau(t)^2-1)$ therefore $\overline\tau(t)=-\coth(\sqrt{-E}(t-c_0))$ for $c_0>0$.  Using these functions and (4.6)-(4.7), we obtain
\begin{equation}a(t)=(-E)^{1/12}e^{-\frac{1}{3}(\lambda+\mu)t}\mbox{csch}^{1/3}(\sqrt{-E}(t-c_0))\end{equation}
\begin{equation}b(t)=C_1(-E)^{1/12}e^{-\frac{1}{3}(\mu-2\lambda)t}\mbox{csch}^{1/3}(\sqrt{-E}(t-c_0))\end{equation}
\begin{equation}c(t)=C_2(-E)^{1/12}e^{-\frac{1}{3}(\lambda-2\mu)t}\mbox{csch}^{1/3}(\sqrt{-E}(t-c_0))\end{equation} for $t>c_0$.  Again since $P(x)=0$, $\phi(t)=\psi _0\equiv$ any constant and $V$ is also a constant given by (5.5) which yields $V=\frac{\sqrt{-E}}{3K^2C_1^2C_2^2}$.


\begin{thebibliography}{99}\setlength{\itemsep}{-2mm}
\bibitem{1} T. Christodoulakis, T. Grammenos, C. Helias, P. Kevrekidis,  G. Papadopoulos, F. Williams, Nova Science Pub. (2005).
\bibitem{1.5}T. Christodoulakis, T. Grammenos, C. Helias, P. Kevrekidis, A. Spanou, {\it J. Math. Phys.}, {\bf 47} (2006) 042505.
  \bibitem{2} J. D'Ambroise, {\it Internat. J. of Pure and Applied Math.}, {\bf 42} (2008), No. 3, p. 405.
  %3+1 FRLW SCH:
\bibitem{3} J. DÕAmbroise, F.L. Williams, {\it Internat. J. of Pure and Applied Math.}, {\bf 34} (2006), No. 1, 117.
%3+1 FRLW EMP with SCH note:
\bibitem{4} R. Hawkins, J. Lidsey, {\it Phys. Rev.}, {\bf D66} (2002), 023523.
%2+1 EMP
\bibitem{5} P. Kevrekidis, F. Williams, {\it Class. Quantum Gravity}, {\bf 20} (2003) L177.
%\bibitem{6}J. Lidsey, Multiple and anisotropic inflation with exponential potentials, {\it Class. Quantum Grav.}, {\bf 9} (1992), 1239-1253.

\bibitem{6} F. Williams, {\it Internat. J. of Modern Physics}, {\bf 20} (2005) 2481.
  \bibitem{7} F. Williams, Proceedings of 11th Marcel Grossmann Meeting, Berlin, 2006.
\end{thebibliography}
\end{document}